\documentclass[12pt, draftclsnofoot, onecolumn]{IEEEtran}

\usepackage{amsfonts}
\usepackage{amsmath}
\usepackage{amssymb}
\usepackage{bbding}
\usepackage{graphicx}
\usepackage{subfigure}
\usepackage{algorithm, algorithmic}
\usepackage{flushend}
\usepackage{booktabs}
\usepackage{multirow}
\usepackage{diagbox}
\usepackage{color}
\usepackage{cite}
\usepackage{hyperref}
\usepackage{cleveref}
\usepackage{stfloats}
\usepackage[table]{xcolor}

\begin{document}

\title{Downlink Precoding for Cell-free FBMC/OQAM Systems With Asynchronous Reception}

\author{Yuhao~Qi,~\IEEEmembership{Student~Member,~IEEE,}
  Jian~Dang,~\IEEEmembership{Senior~Member,~IEEE,}
  Zaichen~Zhang,~\IEEEmembership{Senior~Member,~IEEE,}
  Liang~Wu,~\IEEEmembership{Senior~Member,~IEEE,}
  and~Yongpeng~Wu,~\IEEEmembership{Senior~Member,~IEEE}
\thanks{This work is supported by the National
Key R$\&$D Program of China (2018YFB1801101, 2018YFB1801102), NSFC projects (61971136, 61960206005, 62122052, 62071289), Jiangsu NSF projects (BK20200820, BK20200393), the Fundamental Research Funds for the Central Universities (2242021R41149, 2242022k60001, 3204002302C4), Zhishan Youth Scholar Program of SEU, 111 project (BP0719010), STCSM (22DZ2229005), and Postgraduate Research $\&$ Practice Innovation Program of Jiangsu Province (5004002206).}
\thanks{Y. Qi, J. Dang, Z. Zhang, and L. Wu are with the National Mobile Communications Research Laboratory, Frontiers Science Center for Mobile Information Communication and Security, Southeast University, Nanjing, 210096, China. J. Dang, Z. Zhang and L. Wu are also with the Purple Mountain Laboratory, Nanjing 211111, China. Y. Wu is with the Department of Electronic Engineering, Shanghai Jiao Tong University, Minhang 200240, China (e-mail: qiyuhao@seu.edu.cn; dangjian@seu.edu.cn; zczhang@seu.edu.cn; wuliang@seu.edu.cn; yongpeng.wu@sjtu.edu.cn).}
\thanks{Corresponding author: J. Dang (dangjian@seu.edu.cn)}}

\maketitle

\begin{abstract}
    In this work, an efficient precoding design scheme is proposed for downlink cell-free distributed massive multiple-input multiple-output (DM-MIMO) filter bank multi-carrier (FBMC) systems with asynchronous reception and highly frequency selectivity. The proposed scheme includes a multiple interpolation structure to eliminate the impact of response difference we recently discovered, which has better performance in highly frequency-selective channels. Besides, we also consider the phase shift in asynchronous reception and introduce a phase compensation in the design process. The phase compensation also benefits from the multiple interpolation structure and better adapts to asynchronous reception. Based on the proposed scheme, we theoretically analyze its ergodic achievable rate performance and derive a closed-form expression. Simulation results show that the derived expression can accurately characterize the rate performance, and FBMC with the proposed scheme outperforms orthogonal frequency-division multiplexing (OFDM) in the asynchronous scenario.
\end{abstract}

\begin{IEEEkeywords}
Cell-free DM-MIMO, FBMC, downlink precoding, asynchronous reception, strong frequency selectivity.
\end{IEEEkeywords}

\IEEEpeerreviewmaketitle

\section{Introduction}
The architecture of mobile communication system is undergoing a new transformation with the introduction of cell-free systems \cite{HAHE17}. These systems are based on the principles of distributed massive multiple-input multiple-output (DM-MIMO), where several access points (APs) are linked to a central processing unit (CPU) via optical fibers and simultaneously send signals to all users in a wide area. Since APs are randomly distributed, the distances from each AP to a certain user are different, which causes the signal sent by the nearest AP to the user arriving first, while signals arriving later generates time offsets \cite{LLZW21}. This asynchronous reception degrades the communication system performance, especially in orthogonal frequency-division multiplexing (OFDM) systems, where accurate symbol time synchronization is required to remove cyclic prefix (CP) and perform demodulation.

As a more generalized waveform of OFDM, filter bank multi-carrier (FBMC) \cite{BFB11,RNSM17} has shown robustness in asynchronous scenarios. It consists of time-frequency well localized subcarrier filters, and each symbol stream is processed individually by the filter in that subcarrier, which inhibits high out-of-band radiation, while maintaining symbol orthogonality \cite{RNSM17}. In \cite{YMDD11}, authors provided theoretical performance evaluation of downlink asynchronous multi-carrier systems and demonstrated that FBMC was less sensitive to timing offset due to the better time-frequency localization of the prototype filter. The distortion power for uplink FBMC-based multiple access channel was analysed in \cite{DGXM16}, and the work also showed that FBMC was superior to classic CP-OFDM in the case of asynchronous users. Besides, an iterative receiver was proposed in \cite{SYDA21} for FBMC-based grant-free multiple access system to eliminate the interference caused by timing offsets, which illustrated that FBMC was robust in asynchronous transmissions.

Note that OFDM needs relatively long CP to overcome the impact of both frequency-selective channels and time offset caused by downlink asynchronous reception, which significantly decreases spectral efficiency. On the other hand, due to the absence of CP, the spectral efficiency of FBMC is higher compared to OFDM. A natural question is, whether we can introduce FBMC that is more robust to time offset into the cell-free communication systems to improve the system performance without loss of spectral efficiency? This motivates us to conduct relevant research. In this work, we consider the most popular form of FBMC, i.e., offset quadrature amplitude modulation (FBMC/OQAM) \cite{BFB11}, and the main contributions are summarized as follows:

\begin{enumerate}
    \item We propose an efficient precoding scheme for downlink cell-free DM-MIMO FBMC systems with asynchronous reception and highly frequency-selective channels. Different from conventional multi-tap precoding, the proposed scheme has a multiple interpolation structure, which can better reduce the impact of frequency selectivity and time offset brought by asynchronous reception. 
    \item We analyze the theoretical performance of the proposed scheme and derive the closed-form expression of the achievable rate at receiver, which is not based on the assumption about the flatness of channel response at each subcarrier.
    \item Simulation results show that the derived expression accurately characterizes the rate performance, and FBMC with the proposed scheme outperforms CP-OFDM and FBMC with conventional multi-tap precoding schemes in terms of asynchronous reception.
\end{enumerate}

\emph{Remark}: 
Our previous work \cite{QDZW23} focused on the centralized massive MIMO system, where all receiving antennas were equipped on BS for uplink transmission, and asynchronous reception was not taken into account, which is the main difference compared with this study.

\emph{Notation}: 
$(\cdot)^{*}, (\cdot)^{\top}$ and $(\cdot)^{\mathsf{H}}$ represent the conjugation, transpose and Hermitian transpose, respectively. $\Re\left\{\cdot\right\}$ and $\Im\left\{\cdot\right\}$ are the real and imaginary parts of a complex number. $\mathbb{E}\left\{\cdot\right\}$ denotes the expectation. $\text{diag}\left(\mathbf{y}\right)$ and $\text{vec}\left(\mathbf{Y}\right)$ return a diagonal matrix and column vector given by vector $\mathbf{y}$ and matrix $\mathbf{Y}$, respectively. $\text{Tr}(\cdot)$ denotes the trace. $\star$ and $\otimes$ are linear convolution and kronecker product. $\jmath \triangleq \sqrt{-1}$. $\mathbf{0}$ and $\mathbf{I}_N$ are zero vector and $N\times N$ identity matrix.


\section{System Model}
In this section, we consider a cell-free DM-MIMO FBMC/OQAM downlink system with $U$ single antenna users and $K$ APs each equipped with $N$ antennas. We assume that the total number of subcarriers is $M$, and all users and APs share the same time-frequency resources. In FBMC principle, the input QAM symbols are divided into real and imaginary parts to generate OQAM symbols\cite{BFB11}. Denote the duration of each QAM symbol as $T$, while that of each OQAM symbol is $\frac{T}{2}$. The transmit OQAM symbol vector for all users at the $m$-th subcarrier and $i$-th time index is denoted as $\mathbf{a}_{m,i} \in \mathbb{R}^{U}$, and the symbols are assumed to be independent with the normalized power such that $\mathbb{E}\left\{ \mathbf{a}_{m,i} \mathbf{a}_{m,i}^{\mathsf{H}} \right\} = \mathbf{I}_{U}$. Thus, the synthesized signal of the $k$-th AP is
\begin{align}
    \mathbf{x}_{k}[\ell] = \sum_{i=-\infty}^{\infty} \sum_{m=0}^{M-1} f_{m}\left[\ell-\frac{iM}{2}\right] \mathbf{s}_{m}^{k}[i],
\end{align}
where $f_m[\ell] \triangleq f[\ell] e^{\jmath \frac{2\pi m\ell}{M}}$ is the synthesis filter at the $m$-th subcarrier, and $f[\ell]$ denotes the prototype filter with length $\kappa M$. $\kappa$ represents overlapping factor, and $\ell$ is the sample index corresponding to the sampling interval $T_{\rm s} = \frac{T}{M}$. The signal after precoding is defined as 
\begin{align}
    \mathbf{s}_{m}^{k}[i] \triangleq \sum_{j=-L_{\bar{\rm p}}}^{L_{\bar{\rm p}}} \mathbf{P}_{m}^{k}[j] \mathbf{a}_{m,i-j} e^{\jmath \phi_{m,i-j}}.
\end{align}
By adding the term $\phi_{m,i} = \frac{\pi}{2}(m+i)$, there is a $\pm \frac{\pi}{2}$ phase difference for each OQAM symbol with its neighbors in both time and frequency domains, which can effectively prevent inter-symbol interference (ISI) and inter-carrier interference (ICI). As illustrated in \cite{HABF22}, only single-tap precoder leads to a performance limitation. Thus, in this work, we focus on multi-tap precoders, and $\mathbf{P}_{m}^{k}[i] \in \mathbb{C}^{N\times U}$ represents the precoding matrix of the $i$-th tap with $i$ ranges from $-L_{\bar{\rm p}}$ to $L_{\bar{\rm p}}$.

Considering asynchronous reception effects, the signal received by the $u$-th user from all APs is
\begin{align}\label{eq:y^u}
    y_{u}[\ell] = \sum_{k=1}^{K} \mathbf{h}_{k,u}^{\top}[\ell - \tau_{k,u}] \star \mathbf{x}_{k}[\ell] + \eta_{u}[\ell],
\end{align}
where $\mathbf{h}_{k,u}[\ell] \in \mathbb{C}^{N}$ is the channel vector between the $u$-th user and $k$-th AP. Specifically, it can be expressed as $\mathbf{h}_{k,u}[\ell] = \beta_{k,u}^{1/2} \mathbf{g}_{k,u}[\ell]$, where $\beta_{k,u}$ is large-scale fading coefficient related to the $u$-th user and $k$-th AP, and $\mathbf{g}_{k,u}[\ell]$ denotes small-scale fading with distribution $\mathcal{CN}\left(\mathbf{0}, \lambda_{u}[\ell] \mathbf{I}_{N}\right)$. Here, the sequence $\lambda_{u}[\ell]$ represents the power delay profile (PDP) of $\mathbf{h}_{k,u}[\ell]$, and we assume that the channel responses related to a particular user and different AP antennas are subject to the same channel PDP. Besides, different channel taps are assumed to be independent, and the channels corresponding to different users, APs and AP antennas are also independent. $\tau_{k,u}$ is the time offset intending for the $u$-th user and transmitted by the $k$-th AP. We assume that the start time for each user is the moment when the signal from the nearest AP is received. Thus, $\tau_{k,u}$ can be calculated as $\tau_{k,u} = \frac{d_{k,u} - d_{0,u}}{\mathcal{C} T_{\rm s}}$, where $d_{k,u}$ and $d_{0,u}$ represent the distance from the $k$-th AP and the nearest AP to user-$u$, respectively. $\mathcal{C}$ denotes the speed of light. Note that the unitless $\tau_{k,u}$ here is normalized to the sampling interval, and we only take integer part for the convenience of analysis. $\eta_{u}[\ell]$ represents additive white Gaussian noise (AWGN) and obeys distribution $\mathcal{CN}\left(0, \sigma_{\eta}^{2}\right)$. The $u$-th user recovers its desired OQAM symbols by following normal FBMC/OQAM demodulation steps without any other equalization or multi-user interference cancellation.

\section{Proposed Precoding Design and Performance Analysis}\label{MCE}
In this section, we first demonstrate the proposed precoding design scheme for the downlink cell-free DM-MIMO FBMC systems to overcome the impact brought by asynchronous reception and strong frequency selectivity. Then we evaluate the theoretical rate performance of the proposed scheme.

\subsection{Precoding Design}
Since the precoder works before interpolation, to be more intuitive, we first equate $\mathbf{P}_{m}^{k}[i]$ to the corresponding precoder working after interpolation, i.e., $\mathbf{Q}_{m}^{k}[\ell] = \mathbf{P}_{m}^{k}[i]$ if $\ell = \frac{iM}{2}$ and $\mathbf{0}$ otherwise. In frequency domain, it can be readily found that $\tilde{\mathbf{Q}}_{m}^{k}(\omega) = \tilde{\mathbf{P}}_{m}^{k}\left(\frac{\omega M}{2}\right)$, where $\tilde{\mathbf{Q}}_{m}^{k}(\omega) \triangleq \sum_{\ell} \mathbf{Q}_{m}^{k}[\ell] e^{-\jmath \omega \ell}$ and $\tilde{\mathbf{P}}_{m}^{k}(\omega) \triangleq \sum_{i} \mathbf{P}_{m}^{k}[i] e^{-\jmath \omega i}$ are the responses of $\mathbf{Q}_{m}^{k}[\ell]$ and $\mathbf{P}_{m}^{k}[i]$, respectively. Thus, $\tilde{\mathbf{Q}}_{m}^{k}(\omega)$ is periodic with length $\frac{4\pi}{M}$, and for $\forall m$-th subcarrier with range $\mathcal{B}_{m} = \left[ \frac{2\pi(m-1)}{M}, \frac{2\pi(m+1)}{M} \right]$, $\tilde{\mathbf{Q}}_{m}^{k}\left(\omega_{m}^{\rm L}\right) = \tilde{\mathbf{Q}}_{m}^{k}\left(\omega_{m}^{\rm U}\right)$, where $\omega_{m}^{\rm L}$ and $\omega_{m}^{\rm U}$ represent the lower and upper endpoints of $\mathcal{B}_{m}$. Notice that the time offset $\tau_{k,u}$ in \eqref{eq:y^u} introduces a phase shift in the transmit signal, which has a serious impact on taking real part operation at receivers and thus severely degrades system performance. Thus, in the process of designing precoders, we take the phase shift into consideration and conduct a phase compensation, which yields the goal of precoder design:
\begin{align}\label{eq:HQ}
    \tilde{\mathbf{H}}_{k}(\omega) \tilde{\mathbf{Q}}_{m}^{k}(\omega) = \mathbf{\Lambda}_{k}(\omega), \quad \omega \in \mathcal{B}_{m},
\end{align}
where $\tilde{\mathbf{H}}_{k}(\omega) \triangleq \sum_{\ell} \mathbf{H}_{k}[\ell] e^{-\jmath \omega \ell}$ is the frequency response of $\mathbf{H}_{k}[\ell]$, and the $u$-th row of $\mathbf{H}_{k}[\ell]$ is given by $\mathbf{h}_{k,u}^{\top}[\ell]$. $\mathbf{\Lambda}_{k}(\omega) \triangleq \text{diag}\left( e^{\jmath \omega \tau_{k,1}}, \cdots, e^{\jmath \omega \tau_{k,U}} \right)$. However, $\mathbf{H}_{k}[\ell]$ is random matrix, which yields that $\tilde{\mathbf{H}}_{k}(\omega)$ is also random. Thus, $\tilde{\mathbf{H}}_{k}\left(\omega_{m}^{\rm L}\right) = \tilde{\mathbf{H}}_{k}\left(\omega_{m}^{\rm U}\right)$ is not satisfied with a high probability, which means that \eqref{eq:HQ} is almost impossible to be satisfied at both $\omega_{m}^{\rm L}$ and $\omega_{m}^{\rm U}$ simultaneously. This degrades the performance of the precoder.

\begin{figure}[h]
    \centering
    \includegraphics[width=3in]{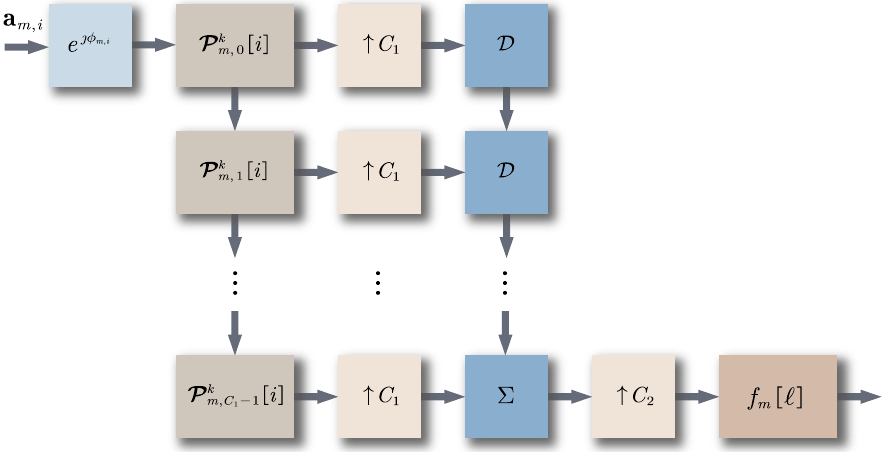}
    \centering
    \caption{Block diagram of the proposed multiple interpolation structure at the $m$-th subcarrier of the $k$-th AP.}
    \label{fig:transmitter}
\end{figure}

Though the periodicity of $\tilde{\mathbf{Q}}_{m}^{k}(\omega)$ is inherent, we can expand the period length so that $\tilde{\mathbf{Q}}_{m}^{k}\left(\omega_{m}^{\rm L}\right) = \tilde{\mathbf{Q}}_{m}^{k}\left(\omega_{m}^{\rm U}\right)$ is not necessary. In this case, $\tilde{\mathbf{Q}}_{m}^{k}(\omega)$ within the range $\mathcal{B}_{m}$ can be carefully designed to better adapt to \eqref{eq:HQ}. In time domain, the expansion of the period length is equivalent to decrease the interpolation factor \cite{QDZW23}. Therefore, another interpolation factor is needed to maintain the total interpolation factor as $\frac{M}{2}$. The proposed structure is shown in Fig. \ref{fig:transmitter}, where $\mathcal{D}$ denotes delay operator, and $C_1$, $C_2$ are the two interpolation factors with $C_1C_2 = \frac{M}{2}$. Since the number of subcarriers is generally taken to the integer power of $2$, $C_1$ and $C_2$ are also taken to the integer power of $2$. Besides, $\boldsymbol{\mathcal{P}}_{m,p}^{k}[i] = \mathbf{P}_{m}^{k}[iC_1+p]$ is a part of the precoder \cite{PPV93}. Notice that the phase compensation also works within the multiple interpolation structure, which is different from the conventional one and can significantly reduce the impact of asynchronous reception.

For the calculation of the precoding coefficients, we assume that each AP has perfect knowledge about channel state information and time offsets between all $U$ users due to space limitation. \eqref{eq:HQ} can be rewritten as $\tilde{\mathbf{H}}_{m,p}^{k} \tilde{\mathbf{Q}}_{m,p}^{k} = \mathbf{\Lambda}_{m,p}^{k}$, where $\tilde{\mathbf{H}}_{m,p}^{k}$, $\tilde{\mathbf{Q}}_{m,p}^{k}$, $\mathbf{\Lambda}_{m,p}^{k}$ represent the values of $\tilde{\mathbf{H}}_{k}(\omega)$, $\tilde{\mathbf{Q}}_{k}(\omega)$, $\mathbf{\Lambda}_{k}(\omega)$ at target frequency bins $\omega_{m,p} = \frac{2\pi \left( m+p/(L_{\bar{\rm p}}+1) \right)}{M}, p = -L_{\bar{\rm p}}, \cdots, L_{\bar{\rm p}}$. In time domain, $\mathbf{P}_{m}^{k}[i]$ can be obtained as
\begin{align}
    \mathbf{P}_{m}^{k} = \left( \mathbf{\Theta}_{m}^{-1} \otimes \mathbf{I}_{N} \right) \mathbf{\Omega}_{m}^{k},
\end{align} 
where the entry of $\mathbf{\Theta}_m \in \mathbb{C}^{L_{\rm p}\times L_{\rm p}}$ in the $(p+L_{\bar{\rm p}})$-th row and $(i+L_{\bar{\rm p}})$-th column is $e^{-\jmath \omega_{m,p} C_{2}i}$. $\mathbf{P}_{m}^{k}, \mathbf{\Omega}_{m}^{k} \in \mathbb{C}^{NL_{\rm p}\times U}$ are block matrices with the $(i+L_{\bar{\rm p}})$-th and $(p+L_{\bar{\rm p}})$-th sub-blocks given by $\mathbf{P}_{m}^{k}[i]$ and $\mathbf{\Omega}_{m,p}^{k}$, respectively, and $\mathbf{\Omega}_{m,p}^{k}$ represents the linear combiner with the phase compensation. Considering maximum ratio combining (MRC) combiner, it can be expressed as
\begin{align}
    \mathbf{\Omega}_{m,p}^{k} = \tilde{\mathbf{H}}_{m,p}^{k\mathsf{H}} \left(\mathbf{\Phi}_{m,p}^{k}\right)^{-1} \mathbf{\Lambda}_{m,p}^{k},
\end{align}
where $\mathbf{\Phi}_{m,p}^{k}$ is a diagonal matrix whose main diagonal is the same as that of $\tilde{\mathbf{H}}_{m,p}^{k} \tilde{\mathbf{H}}_{m,p}^{k\mathsf{H}}$. On the other hand, considering zero forcing (ZF) combiner, it can be expressed as
\begin{align}
    \mathbf{\Omega}_{m,p}^{k} = \tilde{\mathbf{H}}_{m,p}^{k\mathsf{H}} \left( \tilde{\mathbf{H}}_{m,p}^{k} \tilde{\mathbf{H}}_{m,p}^{k\mathsf{H}} \right)^{-1} \mathbf{\Lambda}_{m,p}^{k}.
\end{align}  

\subsection{Theoretical Rate Performance}\label{sec:PA}


At the $\bar{u}$-th receiver, after processed by analysis filter bank, the estimated symbol at the $\bar{m}$-th subcarrier and $i$-th time index can be expressed as  
\begin{align}
    \nonumber
    \hat{a}_{\bar{m},i}^{\bar{u}} = & \sum_{k=1}^{K} \sum_{j=-\infty}^{\infty} \sum_{m=0}^{M-1} \Re \Big\{ \mathbf{c}_{m}^{k,\bar{u}\top}[j] e^{-\jmath \phi_{\bar{m},j}} 
    \\
    & \times \mathbf{a}_{m,i-j} e^{\jmath \phi_{m,i-j}} \Big\} + \hat{\eta}_{\bar{m},i}^{\bar{u}},
\end{align}
where $\hat{\eta}_{\bar{m},i}^{\bar{u}}$ is the noise contained in the estimated symbol.
\begin{align}
    \mathbf{c}_{m}^{k,\bar{u}}\left[ i \right] = \left( f_{m}\left[ \ell \right] \star \mathbf{v}_{m}^{k,\bar{u}}\left[ \ell -\tau_{k,\bar{u}} \right] \star f_{\bar{m}}^{*}\left[ -\ell \right] \right) _{\downarrow \frac{M}{2}}
\end{align}
is the equivalent channel between the precoder at the $m$-th subcarrier and the analysis filter at the $\bar{m}$-th subcarrier. Here, the subscript $\downarrow C$ denotes decimation with factor $C$, and $\mathbf{v}_{m}^{k,\bar{u}}\left[ \ell \right]$ is defined as
\begin{align}
    \mathbf{v}_{m}^{k,\bar{u}}\left[ \ell \right] \triangleq \sum_{i=-L_{\bar{\rm p}}}^{L_{\bar{\rm p}}} \mathbf{P}_{m}^{k\top}\left[ i \right]  \mathbf{h}_{k,\bar{u}} \left[ \ell -iC_2 \right].
\end{align}
To simplify the following analysis, we are interested in the estimated symbol when $i=0$, and symbols at other time indices are regarded as interference. Specifically, $\hat{a}_{\bar{m},0}^{\bar{u}}$ can be further expressed as
\begin{align}\label{eq:hats}
    \nonumber
    \hat{a}_{\bar{m},0}^{\bar{u}} & = \sum_{k,i,\ell,m,u} \Re\left\{ f_{m,i}[\ell] v_{m}^{k,u}[\ell] \right\} a_{m,-i}^{u} + \hat{\eta}_{\bar{m},0}^{\bar{u}}
    \\
    & = \sum_{k,i,m,u} \dot{\boldsymbol{f}}_{m,i}^{\top} \dot{\mathbf{v}}_{m}^{k,u} a_{m,-i}^{u} + \hat{\eta}_{\bar{m},0}^{\bar{u}},
\end{align}
where $v_{m}^{k,u}[\ell]$ is the $u$-th entry of $\mathbf{v}_{m}^{k,\bar{u}}\left[ \ell -\tau_{k,\bar{u}} \right]$, and $\dot{\mathbf{v}}_{m}^{k,u} = \text{vec}\left( \left[ \Re\left\{ \mathbf{v}_{m}^{k,u} \right\}, \Im\left\{ \mathbf{v}_{m}^{k,u} \right\} \right] \right)$ with the vector $\mathbf{v}_{m}^{k,u}$ given by the sequence $v_{m}^{k,u}[\ell]$. On the other hand, $f_{m,i}[\ell]$ is defined as
\begin{align}
    f_{m,i}[\ell] \triangleq f_{m}\left[ -\ell \right] \star f_{\bar{m}}^{*}\left[\ell - \frac{iM}{2}\right] e^{\jmath\left( \phi_{m,-i}-\phi_{\bar{m},0} \right)},
\end{align}
and $\dot{\boldsymbol{f}}_{m,i} = \text{vec}\left( \left[ \Re\left\{ \boldsymbol{f}_{m,i} \right\}, -\Im\left\{ \boldsymbol{f}_{m,i} \right\} \right] \right)$ with the vector $\boldsymbol{f}_{m,i}$ given by the sequence $f_{m,i}[\ell]$. Since channels from different APs to a certain user are independent of each other, the power of symbol $a_{m,-i}^{u}$ contained in $\hat{a}_{\bar{m},0}^{\bar{u}}$ is 
\begin{align}\label{eq:Emiu}
    \mathcal{E}_{m,i}^{u} = \sum_{k=1}^{K} \dot{\boldsymbol{f}}_{m,i}^{\top} \dot{\mathbf{v}}_{m}^{k,u} \dot{\mathbf{v}}_{m}^{k,u\top} \dot{\boldsymbol{f}}_{m,i},
\end{align}
and the ergodic achievable rate of the downlink system can be calculated as \cite[\emph{Lemma 1}]{ZJWZ14}  
\begin{align}\label{eq:Rmu}
    \nonumber
    \mathcal{R}_{\bar{m}}^{\bar{u}} & = \mathbb{E}\left\{ \log_2\left( 1 + \frac{\mathcal{E}_{\bar{m},0}^{\bar{u}}}{\mathop{\sum}_{m,i,u} \mathcal{E}_{m,i}^{u} - \mathcal{E}_{\bar{m},0}^{\bar{u}} + \sigma_{\eta}^{2}/2} \right) \right\}
    \\
    & \approx \log_2\left( 1 + \frac{\mathbb{E}\left\{\mathcal{E}_{\bar{m},0}^{\bar{u}}\right\}}{\mathbb{E}\left\{ \mathop{\sum}_{m,i,u} \mathcal{E}_{m,i}^{u} - \mathcal{E}_{\bar{m},0}^{\bar{u}} \right\} + \sigma_{\eta}^{2}/2} \right).
\end{align}
The approximation will be more and more accurate as the number of AP antennas increases. To obtain the closed-form expression of the rate performance, we will focus on the calculation of $\mathbb{E}\left\{ \mathcal{E}_{m,i}^{u} \right\}$ in the sequence, which depends on $\mathbf{V}_{m}^{k,u} \triangleq \mathbb{E}\left\{ \dot{\mathbf{v}}_{m}^{k,u} \dot{\mathbf{v}}_{m}^{k,u\top} \right\}$. 

\subsubsection{MRC Combiner}
Note that $\mathbf{V}_{m}^{k,u}$ can be written as
\begin{align}\label{eq:Cmku} 
    \mathbf{V}_{m}^{k,u} & = \frac{1}{2}
    \begin{bmatrix}
        \Re \left\{ \dot{\mathbf{V}}_{m}^{k,u} + \ddot{\mathbf{V}}_{m}^{k,u} \right\}, & \Im \left\{ \ddot{\mathbf{V}}_{m}^{k,u} - \dot{\mathbf{V}}_{m}^{k,u} \right\} \\
        \Im \left\{ \dot{\mathbf{V}}_{m}^{k,u} + \ddot{\mathbf{V}}_{m}^{k,u} \right\}, &	\Re \left\{ \dot{\mathbf{V}}_{m}^{k,u} - \ddot{\mathbf{V}}_{m}^{k,u} \right\} \\
    \end{bmatrix},
\end{align}
where $\dot{\mathbf{V}}_{m}^{k,u} = \mathbb{E}\left\{ \mathbf{v}_{m}^{k,u} \mathbf{v}_{m}^{k,u\mathsf{H}} \right\}$, $\ddot{\mathbf{V}}_{m}^{k,u} = \mathbb{E}\left\{ \mathbf{v}_{m}^{k,u} \mathbf{v}_{m}^{k,u\top} \right\}$ are both block matrices with $L_{\rm p}\times L_{\rm p}$ sub-blocks, and there are only $L_{\rm h}\times L_{\rm h}$ non-zero entries for each sub-block. Thus, the entry of the $(i+L_{\bar{\rm p}},j+L_{\bar{\rm p}})$-th sub-block of $\dot{\mathbf{V}}_{m}^{k,u}$ in the $\ell$-th row and $\ell'$-th column can be calculated as
\begin{align}\label{eq:dotc_MRC_rev2}
    \nonumber
    \dot{v}_{m,ij,\ell\ell'}^{k,u} & = \sum_{p,q} \theta_{m,p}^{i} \theta_{m,q}^{j*} \mathbb{E}\Big\{ \boldsymbol{\omega}_{m,p}^{k,u\top} \mathbf{h}_{k,\bar{u}}[\ell-\tau_{k,\bar{u}}] \mathbf{h}_{k,\bar{u}}^{\mathsf{H}}[\ell'-\tau_{k,\bar{u}}] \boldsymbol{\omega}_{m,q}^{k,u*} \Big\}
    \\
    & = \frac{1}{M^2} \sum_{p,q,\dot{m},\ddot{m}} \theta_{m,p}^{i} \theta_{m,q}^{j*} \mathbb{E}\Big\{ \boldsymbol{\omega}_{m,p}^{k,u\top} \tilde{\mathbf{h}}_{\dot{m}}^{k,\bar{u}} \tilde{\mathbf{h}}_{\ddot{m}}^{k,\bar{u}\mathsf{H}} \boldsymbol{\omega}_{m,q}^{k,u*} \Big\} e^{\jmath\frac{2\pi (\dot{m}\ell-\ddot{m}\ell')}{M}} e^{\jmath\frac{2\pi (\ddot{m}-\dot{m})\tau_{k,\bar{u}}}{M}},
\end{align}
where $\theta_{m,p}^{i}$ is the entry of $\mathbf{\Theta}_{m}^{-1}$ in the $(i+L_{\bar{\rm p}})$-th row and $p$-th column, and $\boldsymbol{\omega}_{m,p}^{k,u}$ is the $u$-th column of $\mathbf{\Omega}_{m,p}^{k}$. $\tilde{\mathbf{h}}_{m}^{k,u}$ is the frequency response of the channel $\mathbf{h}_{k,u}[\ell]$ at the $m$-th subcarrier. To remove the correlation between $\boldsymbol{\omega}_{m,p}^{k,u}$ and $\tilde{\mathbf{h}}_{\dot{m}}^{k,\bar{u}}$, we divide $\tilde{\mathbf{h}}_{\dot{m}}^{k,\bar{u}}$ into two parts, i.e., $\tilde{\mathbf{h}}_{\dot{m}}^{k,\bar{u}} = \mu_{\dot{m},p}^{\bar{u}} \tilde{\mathbf{h}}_{m,p}^{k,\bar{u}} + \hat{\mathbf{h}}_{\dot{m},p}^{k,\bar{u}}$, where $\tilde{\mathbf{h}}_{m,p}^{k,\bar{u}}$ is the frequency response of $\mathbf{h}_{k,\bar{u}}[\ell]$ at $\omega_{m,p}$. The coefficient $\mu_{\dot{m},p}^{u}$ is expressed as $\mu_{\dot{m},p}^{u} = \sum_{\ell} \lambda_{u}[\ell] e^{\jmath(\omega_{m,p}-\omega_{\dot{m},0})\ell}$, and $\hat{\mathbf{h}}_{\dot{m},p}^{k,\bar{u}}$ is independent of $\tilde{\mathbf{h}}_{m,p}^{k,\bar{u}}$. As the number of BS antennas increases, $\boldsymbol{\omega}_{m,p}^{k,u}$ tends to $\frac{e^{\jmath \omega_{m,p} \tau_{k,u}}}{N\beta_{k,u}} \tilde{\mathbf{h}}_{m,p}^{k,u*}$. Thus, the correlation between $\boldsymbol{\omega}_{m,p}^{k,u}$ and $\boldsymbol{\omega}_{m,q}^{k,u}$ can be removed by dividing $\tilde{\mathbf{h}}_{m,q}^{k,u}$ as $\tilde{\mathbf{h}}_{m,q}^{k,u} = \zeta_{pq}^{u} \tilde{\mathbf{h}}_{m,p}^{k,u} + \hat{\mathbf{h}}_{m,q}^{k,u}$, where $\zeta_{pq}^{u} = \sum_{\ell}\lambda_{u}[\ell] e^{\jmath\frac{2\pi(p-q)\ell}{(L_{\bar{\rm p}}+1)M}}$ and $\hat{\mathbf{h}}_{m,q}^{k,u}$ is independent of $\tilde{\mathbf{h}}_{m,p}^{k,u}$. The expectation term in \eqref{eq:dotc_MRC_rev2} can be further calculated as
\begin{align}\label{eq:Ewhhw_MRC_rev2}
    \nonumber
    & \mathbb{E}\left\{ \boldsymbol{\omega}_{m,p}^{k,u\top} \tilde{\mathbf{h}}_{\dot{m}}^{k,\bar{u}} \tilde{\mathbf{h}}_{\ddot{m}}^{k,\bar{u}\mathsf{H}} \boldsymbol{\omega}_{m,q}^{k,u*} \right\} 
    \\
    \nonumber
    & = \frac{e^{\jmath(\omega_{m,p}-\omega_{m,q})\tau_{k,u}}}{N^2 \beta_{k,u}^{2}} \mathbb{E}\left\{ \tilde{\mathbf{h}}_{m,p}^{k,u\mathsf{H}} \tilde{\mathbf{h}}_{\dot{m}}^{k,\bar{u}} \tilde{\mathbf{h}}_{\ddot{m}}^{k,\bar{u}\mathsf{H}} \tilde{\mathbf{h}}_{m,q}^{k,u} \right\}
    \\
    \nonumber
    & = \frac{e^{\jmath(\omega_{m,p}-\omega_{m,q})\tau_{k,u}}}{N^2 \beta_{k,u}^{2}} \mathbb{E}\left\{ \tilde{\mathbf{h}}_{m,p}^{k,u\mathsf{H}} \left( \mu_{\dot{m},p}^{\bar{u}} \tilde{\mathbf{h}}_{m,p}^{k,\bar{u}} + \hat{\mathbf{h}}_{\dot{m},p}^{k,\bar{u}} \right) \left( \mu_{\ddot{m},p}^{\bar{u}} \tilde{\mathbf{h}}_{m,p}^{k,\bar{u}} + \hat{\mathbf{h}}_{\ddot{m},p}^{k,\bar{u}} \right)^{\mathsf{H}} \left( \zeta_{pq}^{u} \tilde{\mathbf{h}}_{m,p}^{k,u} + \hat{\mathbf{h}}_{m,q}^{k,u} \right) \right\}
    \\
    & = \frac{\beta_{k,\bar{u}} e^{\jmath(\omega_{m,p}-\omega_{m,q})\tau_{k,u}}}{\beta_{k,u} N^2} \left( \delta_{u\bar{u}} \mu_{\dot{m},p}^{\bar{u}} \mu_{\ddot{m},q}^{\bar{u}*} N^2 + \zeta_{pq}^{u} \xi_{\ddot{m}\dot{m}}^{\bar{u}} N \right),
\end{align}
where $\delta_{u\bar{u}} = 1$ if $u=\bar{u}$ and $0$ otherwise. By applying \eqref{eq:Ewhhw_MRC_rev2} to \eqref{eq:dotc_MRC_rev2}, $\dot{v}_{m,ij,\ell\ell'}^{k,u}$ can be further calculated as
\begin{align}
    \dot{v}_{m,ij,\ell\ell'}^{k,u} = \frac{\beta_{k,\bar{u}}}{\beta_{k,u}} \sum_{p,q} \theta_{m,p}^{i} \theta_{m,q}^{j*} \delta_{u\bar{u}} \lambda_{m,p}^{k,u}[\ell] \lambda_{m,q}^{k,u*}[\ell'] + \theta_{m,p}^{i} \theta_{m,q}^{j*} \zeta_{pq}^{k,u} \frac{\lambda_{\bar{u}}[\ell-\tau_{k,\bar{u}}]}{N}.
\end{align}
Hence, the $(i+L_{\bar{\rm p}}, j+L_{\bar{\rm p}})$-th sub-block of $\dot{\mathbf{V}}_{m}^{k,u}$ is expressed as
\begin{align}\label{eq:dotC_MRC}
    \dot{\mathbf{V}}_{m,ij}^{k,u} = \frac{\beta_{k,\bar{u}}}{\beta_{k,u}} \sum_{p,q} \theta_{m,p}^{i} \theta_{m,q}^{j*} \delta_{u\bar{u}} \dot{\mathbf{D}}_{m,pq}^{k,u} + \theta_{m,p}^{i} \theta_{m,q}^{j*} \zeta_{pq}^{k,u} \frac{\mathbf{\Lambda}_{k,\bar{u}}}{N},
\end{align}
$\mathbf{\Lambda}_{k,u}$ is a diagonal matrix whose main diagonal is given by $\lambda_{k,u}[\ell] \triangleq \lambda_{u}[\ell-\tau_{k,u}]$. $\zeta_{pq}^{k,u}$ is defined as $\zeta_{pq}^{k,u} \triangleq \zeta_{pq}^{u} e^{\jmath \frac{2\pi (p-q)\tau_{k,u}}{(L_{\bar{\rm p}}+1)M}}$. Besides, $\dot{\mathbf{D}}_{m,pq}^{k,u} = \boldsymbol{\lambda}_{m,p}^{k,u} \boldsymbol{\lambda}_{m,q}^{k,u\mathsf{H}}$ with the entries of column vector $\boldsymbol{\lambda}_{m,p}^{k,u}$ given by $\lambda_{m,p}^{k,u}[\ell] \triangleq \lambda_{k,u}[\ell] e^{\jmath \omega_{m,p} \ell}$. Following the similar way, the entry of the $(i+L_{\bar{\rm p}},j+L_{\bar{\rm p}})$-th sub-block of $\ddot{\mathbf{V}}_{m}^{k,u}$ in the $\ell$-th row and $\ell'$-th column can be calculated as
\begin{align}\label{eq:ddotc_MRC_rev2}
    \nonumber
    \ddot{v}_{m,ij,\ell\ell'}^{k,u} & = \sum_{p,q} \theta_{m,p}^{i} \theta_{m,q}^{j} \mathbb{E}\left\{ \boldsymbol{\omega}_{m,p}^{k,u\top} \mathbf{h}_{k,\bar{u}}[\ell-\tau_{k,\bar{u}}] \boldsymbol{\omega}_{m,q}^{k,u\top} \mathbf{h}_{k,\bar{u}}[\ell'-\tau_{k,\bar{u}}] \right\}
    \\
    & = \frac{1}{M^2} \sum_{p,q,\dot{m},\ddot{m}} \theta_{m,p}^{i} \theta_{m,q}^{j} \mathbb{E}\left\{ \boldsymbol{\omega}_{m,p}^{k,u\top} \tilde{\mathbf{h}}_{\dot{m}}^{k,\bar{u}} \boldsymbol{\omega}_{m,q}^{k,u\top} \tilde{\mathbf{h}}_{\ddot{m}}^{k,\bar{u}} \right\} e^{\jmath\frac{2\pi (\dot{m}\ell+\ddot{m}\ell')}{M}} e^{-\jmath\frac{2\pi (\dot{m}+\ddot{m})\tau_{k,\bar{u}}}{M}}.
\end{align}
The expectation term in \eqref{eq:ddotc_MRC_rev2} can be further calculated as
\begin{align}\label{eq:Ewhwh_MRC_rev2}
    \nonumber
    & \mathbb{E}\left\{ \boldsymbol{\omega}_{m,p}^{k,u\top} \tilde{\mathbf{h}}_{\dot{m}}^{k,\bar{u}} \tilde{\mathbf{h}}_{\ddot{m}}^{k,\bar{u}\mathsf{H}} \boldsymbol{\omega}_{m,q}^{k,u*} \right\} 
    \\
    \nonumber
    & = \frac{e^{\jmath(\omega_{m,p}+\omega_{m,q})\tau_{k,u}}}{N^2 \beta_{k,u}^{2}} \mathbb{E}\left\{ \tilde{\mathbf{h}}_{m,p}^{k,u\mathsf{H}} \tilde{\mathbf{h}}_{\dot{m}}^{k,\bar{u}} \tilde{\mathbf{h}}_{m,q}^{k,u\mathsf{H}} \tilde{\mathbf{h}}_{\ddot{m}}^{k,\bar{u}} \right\}
    \\
    & = \frac{\delta_{u\bar{u}} e^{\jmath(\omega_{m,p}+\omega_{m,q})\tau_{k,u}}}{N^2} \left( \mu_{\dot{m},p}^{\bar{u}} \mu_{\ddot{m},q}^{\bar{u}} N^2 + \mu_{\dot{m},q}^{\bar{u}} \mu_{\ddot{m},p}^{\bar{u}} N \right).
\end{align}
By applying \eqref{eq:Ewhwh_MRC_rev2} to \eqref{eq:ddotc_MRC_rev2}, $\ddot{v}_{m,ij,\ell\ell'}^{k,u}$ can be further calculated as
\begin{align}
    \ddot{v}_{m,ij,\ell\ell'}^{k,u} = \sum_{p,q} \theta_{m,p}^{i} \theta_{m,q}^{j} \delta_{u\bar{u}} \lambda_{m,p}^{k,u}[\ell] \lambda_{m,q}^{k,u}[\ell'] + \theta_{m,p}^{i} \theta_{m,q}^{j} \delta_{u\bar{u}} \frac{\lambda_{m,q}^{k,u}[\ell] \lambda_{m,p}^{k,u}[\ell']}{N}.
\end{align}
Hence, only when $u=\bar{u}$, $\ddot{\mathbf{V}}_{m}^{k,u}$ is a non-zero matrix, and its $(i+L_{\bar{\rm p}}, j+L_{\bar{\rm p}})$-th sub-block is
\begin{align}\label{eq:ddotC_MRC}
    \ddot{\mathbf{V}}_{m,ij}^{k,u} = \sum_{p,q} \theta_{m,p}^{i} \theta_{m,q}^{j} \ddot{\mathbf{D}}_{m,pq}^{k,u} + \theta_{m,p}^{i} \theta_{m,q}^{j} \frac{\ddot{\mathbf{D}}_{m,qp}^{k,\bar{u}}}{N},
\end{align}
where $\ddot{\mathbf{D}}_{m,pq}^{k,u} = \boldsymbol{\lambda}_{m,p}^{k,u} \boldsymbol{\lambda}_{m,q}^{k,u\top}$. By applying \eqref{eq:dotC_MRC}, \eqref{eq:ddotC_MRC} into \eqref{eq:Cmku} and integrating equations \eqref{eq:Emiu}, \eqref{eq:Rmu}, \eqref{eq:Cmku}, the closed-form expression of $\mathcal{R}_{\bar{m}}^{\bar{u}}$ with MRC combiner can be obtained.

\subsubsection{ZF Combiner}
In this case, the entry of the $(i+L_{\bar{\rm p}},j+L_{\bar{\rm p}})$-th sub-block of $\dot{\mathbf{V}}_{m}^{k,u}$ in the $\ell$-th row and $\ell'$-th column can be calculated as
\begin{align}\label{eq:dotc_rev2}
    \nonumber
    & \dot{v}_{m,ij,\ell\ell'}^{k,u} = \sum_{p,q} \theta_{m,p}^{i} \theta_{m,q}^{j*} \mathbb{E}\left\{ \boldsymbol{\omega}_{m,p}^{k,u\top} \mathbf{h}_{k,\bar{u}}[\ell-\tau_{k,\bar{u}}] \mathbf{h}_{k,\bar{u}}^{\mathsf{H}}[\ell'-\tau_{k,\bar{u}}] \boldsymbol{\omega}_{m,q}^{k,u*} \right\}
    \\
    \nonumber
    & = \frac{1}{M^2} \sum_{p,q,\dot{m},\ddot{m}} \theta_{m,p}^{i} \theta_{m,q}^{j*} \mathbb{E}\left\{ \boldsymbol{\omega}_{m,p}^{k,u\top} \tilde{\mathbf{h}}_{\dot{m}}^{k,\bar{u}} \tilde{\mathbf{h}}_{\ddot{m}}^{k,\bar{u}\mathsf{H}} \boldsymbol{\omega}_{m,q}^{k,u*} \right\} e^{\jmath\frac{2\pi (\dot{m}\ell-\ddot{m}\ell')}{M}} e^{\jmath\frac{2\pi (\ddot{m}-\dot{m})\tau_{k,\bar{u}}}{M}}
    \\
    \nonumber
    & = \frac{1}{M^2} \sum_{p,q,\dot{m},\ddot{m}} \theta_{m,p}^{i} \theta_{m,q}^{j*} \mathbb{E}\left\{ \boldsymbol{\omega}_{m,p}^{k,u\top} \left( \mu_{\dot{m},p}^{\bar{u}} \tilde{\mathbf{h}}_{m,p}^{k,\bar{u}} + \hat{\mathbf{h}}_{\dot{m},p}^{k,\bar{u}} \right) \left( \mu_{\ddot{m},q}^{\bar{u}} \tilde{\mathbf{h}}_{m,q}^{k,\bar{u}} + \hat{\mathbf{h}}_{\ddot{m},q}^{k,\bar{u}} \right)^{\mathsf{H}} \boldsymbol{\omega}_{m,q}^{k,u*} \right\}
    \\
    \nonumber
    & \quad \times e^{\jmath\frac{2\pi (\dot{m}\ell-\ddot{m}\ell')}{M}} e^{\jmath\frac{2\pi (\ddot{m}-\dot{m})\tau_{k,\bar{u}}}{M}}
    \\
    \nonumber
    & \overset{\text{(a)}}{=} \frac{1}{M^2} \sum_{p,q,\dot{m},\ddot{m}} \theta_{m,p}^{i} \theta_{m,q}^{j*} \mathbb{E} \left\{ \boldsymbol{\omega}_{m,p}^{k,u\top} \hat{\mathbf{h}}_{\dot{m},p}^{k,\bar{u}} \hat{\mathbf{h}}_{\ddot{m},q}^{k,\bar{u}\mathsf{H}} \boldsymbol{\omega}_{m,q}^{k,u*} \right\} e^{\jmath\frac{2\pi (\dot{m}\ell-\ddot{m}\ell')}{M}} e^{\jmath\frac{2\pi (\ddot{m}-\dot{m})\tau_{k,\bar{u}}}{M}}
    \\
    \nonumber
    & \quad + \frac{\beta_{k,\bar{u}}}{\beta_{k,u} M^2} \sum_{p,q,\dot{m},\ddot{m}} \delta_{u\bar{u}} \theta_{m,p}^{i} \theta_{m,q}^{j*} \mu_{\dot{m},p}^{\bar{u}} \mu_{\ddot{m},q}^{\bar{u}*} e^{\jmath\frac{2\pi (\dot{m}\ell-\ddot{m}\ell')}{M}} e^{\jmath\frac{2\pi (\ddot{m}-\dot{m})\tau_{k,\bar{u}}}{M}} e^{\jmath(\omega_{m,p}-\omega_{m,q})\tau_{k,\bar{u}}}
    \\
    \nonumber
    & = \frac{1}{M^2} \sum_{p,q,\dot{m},\ddot{m}} \theta_{m,p}^{i} \theta_{m,q}^{j*} \mathbb{E} \left\{ \boldsymbol{\omega}_{m,p}^{k,u\top} \hat{\mathbf{h}}_{\dot{m},p}^{k,\bar{u}} \hat{\mathbf{h}}_{\ddot{m},q}^{k,\bar{u}\mathsf{H}} \boldsymbol{\omega}_{m,q}^{k,u*} \right\} e^{\jmath\frac{2\pi (\dot{m}\ell-\ddot{m}\ell')}{M}} e^{\jmath\frac{2\pi (\ddot{m}-\dot{m})\tau_{k,\bar{u}}}{M}}
    \\
    & \quad + \frac{\beta_{k,\bar{u}}}{\beta_{k,u}} \sum_{p,q} \delta_{u\bar{u}} \theta_{m,p}^{i} \theta_{m,q}^{j*} \lambda_{m,p}^{k,u}[\ell] \lambda_{m,q}^{k,u*}[\ell'].
\end{align}
The equation (a) in \eqref{eq:dotc_rev2} holds due to the fact that $\boldsymbol{\omega}_{m,p}^{k,u\top} \tilde{\mathbf{h}}_{m,p}^{k,\bar{u}} = e^{\jmath \omega_{m,p} \tau_{k,\bar{u}}}$ if $u = \bar{u}$ and $0$ otherwise. Since independence only exists between $\boldsymbol{\omega}_{m,p}^{k,u}$ and $\hat{\mathbf{h}}_{\dot{m},p}^{k,\bar{u}}$ and between $\boldsymbol{\omega}_{m,q}^{k,u}$ and $\hat{\mathbf{h}}_{\ddot{m},q}^{k,\bar{u}}$ with every other two vectors are correlated, it is more difficult to solve the expectation term in \eqref{eq:dotc_rev2}. Moreover, the matrix inverse operations of $\boldsymbol{\omega}_{m,p}^{k,u}$ and $\boldsymbol{\omega}_{m,q}^{k,u}$ further increase the difficulty. To remove the correlation between the vectors, $\tilde{\mathbf{H}}_{m,q}^{k}$ can be rewritten as $\tilde{\mathbf{H}}_{m,q}^{k} = \mathbf{\Lambda}_{\zeta,pq} \tilde{\mathbf{H}}_{m,p}^{k} + \hat{\mathbf{H}}_{m,q}^{k}$, where $\hat{\mathbf{H}}_{m,q}^{k}$ is independent of $\tilde{\mathbf{H}}_{m,p}^{k}$. $\mathbf{\Lambda}_{\zeta,pq} = \text{diag}\left( \zeta_{pq}^{1}, \cdots, \zeta_{pq}^{U} \right)$. Besides, according to Taylor expansion, the inverse for an invertible matrix $\mathbf{X}^{-1} + \mathbf{Y}$ can be expressed as $\left( \mathbf{X}^{-1} + \mathbf{Y} \right)^{-1} = \mathbf{X} - \mathbf{X} \mathbf{Y} \mathbf{X} + \mathbf{X} \left(\mathbf{Y} \mathbf{X}\right)^2 - \mathbf{X} \left(\mathbf{Y} \mathbf{X}\right)^3 + \cdots$. For the convenience of expression, we denote $\check{\mathbf{H}}_{m,p}^{k} \triangleq \mathbf{\Lambda}_{m,p}^{k\mathsf{H}} \tilde{\mathbf{H}}_{m,p}^{k}$. Therefore, $\mathbf{\Omega}_{m,q}^{k}$ can be written as $\mathbf{\Omega}_{m,q}^{k} = \check{\mathbf{H}}_{m,q}^{k\mathsf{H}} \left( \check{\mathbf{H}}_{m,q}^{k} \check{\mathbf{H}}_{m,q}^{k\mathsf{H}} \right)^{-1}$ and approximated by its first order Taylor expansion, i.e.,
\begin{align}\label{eq:Omega_mqk_rev2}
    \nonumber
    \mathbf{\Omega}_{m,q}^{k} & = \left(\mathbf{\Lambda}_{\zeta,pq}^{k} \check{\mathbf{H}}_{m,p}^{k} + \mathbf{\Lambda}_{m,q}^{k\mathsf{H}} \hat{\mathbf{H}}_{m,q}^{k}\right)^{\mathsf{H}} \left( \left(\mathbf{\Lambda}_{\zeta,pq}^{k} \check{\mathbf{H}}_{m,p}^{k} + \mathbf{\Lambda}_{m,q}^{k\mathsf{H}} \hat{\mathbf{H}}_{m,q}^{k}\right) \left(\mathbf{\Lambda}_{\zeta,pq}^{k} \check{\mathbf{H}}_{m,p}^{k} + \mathbf{\Lambda}_{m,q}^{k\mathsf{H}} \hat{\mathbf{H}}_{m,q}^{k}\right)^{\mathsf{H}}\right)^{-1}
    \\
    & \approx \mathbf{\Omega}_{m,p}^{k} \left(\mathbf{\Lambda}_{\zeta,pq}^{k}\right)^{-1} + \hat{\mathbf{H}}_{m,q}^{k\mathsf{H}} \mathbf{\Lambda}_{m,q}^{k} \left( \mathbf{\Lambda}_{\zeta,pq}^{k} \check{\mathbf{H}}_{m,p}^{k} \check{\mathbf{H}}_{m,p}^{k\mathsf{H}} \mathbf{\Lambda}_{\zeta,pq}^{k\mathsf{H}} \right)^{-1},
\end{align}
where $\mathbf{\Lambda}_{\zeta,pq}^{k} = \mathbf{\Lambda}_{m,q}^{k\mathsf{H}} \mathbf{\Lambda}_{\zeta,pq} \mathbf{\Lambda}_{m,p}^{k}$. Note that $\hat{\mathbf{h}}_{\ddot{m},q}^{k,\bar{u}}$ is correlated with $\boldsymbol{\omega}_{m,p}^{k,u}$, it can be transformed into $\hat{\mathbf{h}}_{\ddot{m},q}^{k,\bar{u}} = \left(\mu_{\ddot{m},p}^{\bar{u}} - \mu_{\ddot{m},q}^{\bar{u}} \zeta_{pq}^{\bar{u}}\right) \tilde{\mathbf{h}}_{m,p}^{k,\bar{u}} + \hat{\mathbf{h}}_{\ddot{m},pq}^{k,\bar{u}}$, where $\hat{\mathbf{h}}_{\ddot{m},pq}^{k,\bar{u}}$ is independent of $\boldsymbol{\omega}_{m,p}^{k,u}$. The expectation term in \eqref{eq:dotc_rev2} can be further calculated as
\begin{align}\label{eq:Ewhhw_rev2}
    \nonumber
    & \mathbb{E} \left\{ \boldsymbol{\omega}_{m,p}^{k,u\top} \hat{\mathbf{h}}_{\dot{m},p}^{k,\bar{u}} \hat{\mathbf{h}}_{\ddot{m},q}^{k,\bar{u}\mathsf{H}} \boldsymbol{\omega}_{m,q}^{k,u*} \right\} \approx (\zeta_{qp}^{k,u})^{-1} \mathbb{E}\left\{ \boldsymbol{\omega}_{m,p}^{k,u\top} \hat{\mathbf{h}}_{\dot{m},p}^{k,\bar{u}} \hat{\mathbf{h}}_{\ddot{m},pq}^{k,\bar{u}\mathsf{H}} \boldsymbol{\omega}_{m,p}^{k,u*} \right\}
    \\
    \nonumber
    & = \frac{\mathbb{E}\left\{ \boldsymbol{\omega}_{m,p}^{k,u\top} \mathbb{E}\left\{ \left( \tilde{\mathbf{h}}_{\dot{m}}^{k,\bar{u}} - \mu_{\dot{m},p}^{\bar{u}} \tilde{\mathbf{h}}_{m,p}^{k,\bar{u}} \right) \left( \tilde{\mathbf{h}}_{\ddot{m}}^{k,\bar{u}} - \mu_{\ddot{m},q}^{\bar{u}} \tilde{\mathbf{h}}_{m,q}^{k,\bar{u}} - \left( \mu_{\ddot{m},p}^{\bar{u}} - \mu_{\ddot{m},q}^{\bar{u}} \zeta_{pq}^{\bar{u}} \right) \tilde{\mathbf{h}}_{m,p}^{k,\bar{u}} \right)^{\mathsf{H}} \right\} \boldsymbol{\omega}_{m,p}^{k,u*} \right\}}{\zeta_{qp}^{k,u}}
    \\
    & \overset{\text{(a)}}{=} \frac{\beta_{k,\bar{u}} \left(\xi_{\ddot{m}\dot{m}}^{\bar{u}} - \mu_{\ddot{m},p}^{\bar{u}*} \mu_{\dot{m},p}^{\bar{u}} - \mu_{\ddot{m},q}^{\bar{u}*} \mu_{\dot{m},q}^{\bar{u}} + \mu_{\dot{m},p}^{\bar{u}} \mu_{\ddot{m},q}^{\bar{u}*} \zeta_{qp}^{\bar{u}}\right)}{\beta_{k,u} \zeta_{qp}^{k,u} (N-U)},
\end{align}
where $\xi_{\ddot{m}\dot{m}}^{u} = \sum_{\ell} \lambda_{u}[\ell] e^{\jmath\frac{2\pi(\ddot{m}-\dot{m})\ell}{M}}$. Equation (a) in \eqref{eq:Ewhhw_rev2} holds due to the fact that for a complex Wishart matrix $\mathbf{X}$ with distribution $\mathcal{CW}_{q}(p,\mathbf{I}_q)$, $\mathbb{E}\left\{ \text{Tr}\left( \mathbf{X}^{-1} \right) \right\} = \frac{q}{p-q}$. By applying \eqref{eq:Ewhhw_rev2} to \eqref{eq:dotc_rev2}, $\dot{v}_{m,ij,\ell\ell'}^{k,u}$ can be further calculated as 
\begin{align}
    \nonumber
    & \dot{v}_{m,ij,\ell\ell'}^{k,u} \approx \frac{\beta_{k,\bar{u}}}{\beta_{k,u}}\sum_{p,q} \delta_{u\bar{u}} \theta_{m,p}^{i} \theta_{m,q}^{j*} \lambda_{m,p}^{k,u}[\ell] \lambda_{m,q}^{k,u*}[\ell']
    \\
    & + \theta_{m,p}^{i} \theta_{m,q}^{j*} \frac{\delta_{\ell\ell'} \lambda_{\bar{u}}[\ell-\tau_{k,\bar{u}}] - \lambda_{m,p}^{k,\bar{u}}[\ell] \lambda_{m,p}^{k,\bar{u}*}[\ell'] - \lambda_{m,q}^{k,\bar{u}}[\ell] \lambda_{m,q}^{k,\bar{u}*}[\ell'] + \zeta_{qp}^{k,\bar{u}} \lambda_{m,p}^{k,\bar{u}}[\ell] \lambda_{m,q}^{k,\bar{u}*}[\ell']}{\zeta_{qp}^{k,u}(N-U)}.
\end{align}
Hence, the $(i+L_{\bar{\rm p}}, j+L_{\bar{\rm p}})$-th sub-block of $\dot{\mathbf{V}}_{m}^{k,u}$ can be expressed as
\begin{align}\label{eq:dotCii}
    \dot{\mathbf{V}}_{m,ij}^{k,u} = \frac{\beta_{k,\bar{u}}}{\beta_{k,u}} \sum_{p,q} \theta_{m,p}^{i} \theta_{m,q}^{j*} \delta_{u\bar{u}} \dot{\mathbf{D}}_{m,pq}^{k,u} + \theta_{m,p}^{i} \theta_{m,q}^{j*}
    \frac{\mathbf{\Lambda}_{k,\bar{u}} - \dot{\mathbf{D}}_{m,pp}^{k,\bar{u}} - \dot{\mathbf{D}}_{m,qq}^{k,\bar{u}} + \zeta_{qp}^{k,\bar{u}} \dot{\mathbf{D}}_{m,pq}^{k,\bar{u}}}{\zeta_{qp}^{k,u}(N-U)}.
\end{align}
Similarly, the entry of the $(i+L_{\bar{\rm p}},j+L_{\bar{\rm p}})$-th sub-block of $\ddot{\mathbf{V}}_{m}^{k,u}$ in the $\ell$-th row and $\ell'$-th column can be calculated as
\begin{align}\label{eq:ddotc_rev2}
    \nonumber
    & \ddot{v}_{m,ij,\ell\ell'}^{k,u} = \sum_{p,q} \theta_{m,p}^{i} \theta_{m,q}^{j} \mathbb{E}\left\{ \boldsymbol{\omega}_{m,p}^{k,u\top} \mathbf{h}_{k,\bar{u}}[\ell-\tau_{k,\bar{u}}] \boldsymbol{\omega}_{m,q}^{k,u\top} \mathbf{h}_{k,\bar{u}}[\ell-\tau_{k,\bar{u}}] \right\}
    \\
    \nonumber
    & = \frac{1}{M^2} \sum_{p,q,\dot{m},\ddot{m}} \theta_{m,p}^{i} \theta_{m,q}^{j} \mathbb{E}\left\{ \boldsymbol{\omega}_{m,p}^{k,u\top} \tilde{\mathbf{h}}_{\dot{m}}^{k,\bar{u}} \boldsymbol{\omega}_{m,q}^{k,u\top} \tilde{\mathbf{h}}_{\ddot{m}}^{k,\bar{u}} \right\} e^{\jmath\frac{2\pi (\dot{m}\ell+\ddot{m}\ell')}{M}} e^{-\jmath\frac{2\pi (\dot{m}+\ddot{m})\tau_{k,\bar{u}}}{M}}
    \\
    \nonumber
    & = \frac{1}{M^2} \sum_{p,q,\dot{m},\ddot{m}} \theta_{m,p}^{i} \theta_{m,q}^{j} \mathbb{E}\left\{ \boldsymbol{\omega}_{m,p}^{k,u\top} \left( \mu_{\dot{m},p}^{\bar{u}} \tilde{\mathbf{h}}_{m,p}^{k,\bar{u}} + \hat{\mathbf{h}}_{\dot{m},p}^{k,\bar{u}} \right) \boldsymbol{\omega}_{m,q}^{k,u\top} \left( \mu_{\ddot{m},q}^{\bar{u}} \tilde{\mathbf{h}}_{m,q}^{k,\bar{u}} + \hat{\mathbf{h}}_{\ddot{m},q}^{k,\bar{u}} \right) \right\}
    \\
    \nonumber
    & \quad \times e^{\jmath\frac{2\pi (\dot{m}\ell+\ddot{m}\ell')}{M}} e^{-\jmath\frac{2\pi (\dot{m}+\ddot{m})\tau_{k,\bar{u}}}{M}}
    \\
    \nonumber
    & = \frac{1}{M^2} \sum_{p,q,\dot{m},\ddot{m}} \theta_{m,p}^{i} \theta_{m,q}^{j} \mathbb{E} \left\{ \boldsymbol{\omega}_{m,p}^{k,u\top} \hat{\mathbf{h}}_{\dot{m},p}^{k,\bar{u}} \boldsymbol{\omega}_{m,q}^{k,u\top} \hat{\mathbf{h}}_{\ddot{m},q}^{k,\bar{u}} \right\} e^{\jmath\frac{2\pi (\dot{m}\ell+\ddot{m}\ell')}{M}} e^{-\jmath\frac{2\pi (\dot{m}+\ddot{m})\tau_{k,\bar{u}}}{M}}
    \\
    \nonumber
    & \quad + \frac{\beta_{k,\bar{u}}}{\beta_{k,u} M^2} \sum_{p,q,\dot{m},\ddot{m}} \delta_{u\bar{u}} \theta_{m,p}^{i} \theta_{m,q}^{j} \mu_{\dot{m},p}^{\bar{u}} \mu_{\ddot{m},q}^{\bar{u}} e^{\jmath\frac{2\pi (\dot{m}\ell+\ddot{m}\ell')}{M}} e^{-\jmath\frac{2\pi (\dot{m}+\ddot{m})\tau_{k,\bar{u}}}{M}} e^{\jmath(\omega_{m,p}+\omega_{m,q})\tau_{k,\bar{u}}}
    \\
    \nonumber
    & = \frac{1}{M^2} \sum_{p,q,\dot{m},\ddot{m}} \theta_{m,p}^{i} \theta_{m,q}^{j} \mathbb{E} \left\{ \boldsymbol{\omega}_{m,p}^{k,u\top} \hat{\mathbf{h}}_{\dot{m},p}^{k,\bar{u}} \boldsymbol{\omega}_{m,q}^{k,u\top} \hat{\mathbf{h}}_{\ddot{m},q}^{k,\bar{u}} \right\} e^{\jmath\frac{2\pi (\dot{m}\ell+\ddot{m}\ell')}{M}} e^{-\jmath\frac{2\pi (\dot{m}+\ddot{m})\tau_{k,\bar{u}}}{M}}
    \\
    & \quad + \frac{\beta_{k,\bar{u}}}{\beta_{k,u}} \sum_{p,q} \delta_{u\bar{u}} \theta_{m,p}^{i} \theta_{m,q}^{j} \lambda_{m,p}^{k,u}[\ell] \lambda_{m,q}^{k,u}[\ell'].
\end{align}
According to \eqref{eq:Omega_mqk_rev2}, the expectation term in \eqref{eq:ddotc_rev2} is approximated as
\begin{align}\label{eq:Ewhwh_rev2}
    \nonumber
    & \mathbb{E} \left\{ \boldsymbol{\omega}_{m,p}^{k,u\top} \hat{\mathbf{h}}_{\dot{m},p}^{k,\bar{u}} \boldsymbol{\omega}_{m,q}^{k,u\top} \hat{\mathbf{h}}_{\ddot{m},q}^{k,\bar{u}} \right\}
    \\
    \nonumber
    & \approx \left(\mu_{\ddot{m},p}^{\bar{u}} - \mu_{\ddot{m},q}^{\bar{u}} \zeta_{pq}^{\bar{u}}\right) \mathbb{E}\left\{ \boldsymbol{\omega}_{m,p}^{k,u\top} \hat{\mathbf{h}}_{\dot{m},p}^{k,\bar{u}} \left(\hat{\mathbf{H}}_{m,q}^{k\mathsf{H}} \mathbf{\Lambda}_{m,q}^{k} \left( \mathbf{\Lambda}_{\zeta,pq}^{k} \check{\mathbf{H}}_{m,p}^{k} \check{\mathbf{H}}_{m,p}^{k\mathsf{H}} \mathbf{\Lambda}_{\zeta,pq}^{k\mathsf{H}} \right)^{-1} \mathbf{e}_{u} \right)^{\top} \tilde{\mathbf{h}}_{m,p}^{k,\bar{u}} \right\}
    \\
    \nonumber
    & = \frac{\delta_{u\bar{u}} \beta_{k,\bar{u}} e^{\jmath\omega_{m,p}\tau_{k,\bar{u}}} \left( \mu_{\ddot{m},p}^{\bar{u}} - \mu_{\ddot{m},q}^{\bar{u}} \zeta_{pq}^{\bar{u}}\right)\left( \mu_{\dot{m},q}^{\bar{u}} - \mu_{\dot{m},p}^{\bar{u}} \zeta_{qp}^{\bar{u}} \right)}{\big|\zeta_{pq}^{k,\bar{u}}\big|^2} \mathbb{E}\left\{ \left( \mathbf{e}_{\bar{u}}^{\top} \mathbf{\Lambda}_{m,q}^{k} \left( \check{\mathbf{H}}_{m,p}^{k} \check{\mathbf{H}}_{m,p}^{k\mathsf{H}} \right)^{-1} \mathbf{e}_{u} \right)^{\top} \right\}
    \\
    & = \frac{\delta_{u\bar{u}} e^{\jmath(\omega_{m,p}+\omega_{m,q})\tau_{k,\bar{u}}} \left( \mu_{\ddot{m},p}^{\bar{u}} - \mu_{\ddot{m},q}^{\bar{u}} \zeta_{pq}^{\bar{u}}\right)\left( \mu_{\dot{m},q}^{\bar{u}} - \mu_{\dot{m},p}^{\bar{u}} \zeta_{qp}^{\bar{u}} \right)}{\big|\zeta_{pq}^{k,\bar{u}}\big|^2 (N-U)}.
\end{align}
By applying \eqref{eq:Ewhwh_rev2} to \eqref{eq:ddotc_rev2}, $\ddot{v}_{m,ij,\ell\ell'}^{k,u}$ can be further calculated as
\begin{align}
    \nonumber
    & \ddot{v}_{m,ij,\ell\ell'}^{k,u} \approx \sum_{p,q} \delta_{u\bar{u}} \theta_{m,p}^{i} \theta_{m,q}^{j} \lambda_{m,p}^{k,u}[\ell] \lambda_{m,q}^{k,u}[\ell']
    \\
    & + \delta_{u\bar{u}} \theta_{m,p}^{i} \theta_{m,q}^{j} \frac{ \lambda_{m,p}^{k,\bar{u}}[\ell'] \lambda_{m,q}^{k,\bar{u}}[\ell] - \zeta_{qp}^{k,\bar{u}} \lambda_{m,p}^{k,\bar{u}}[\ell] \lambda_{m,p}^{k,\bar{u}}[\ell'] - \zeta_{pq}^{k,\bar{u}} \lambda_{m,q}^{k,\bar{u}}[\ell] \lambda_{m,q}^{k,\bar{u}}[\ell'] + \big|\zeta_{qp}^{k,\bar{u}}\big|^2 \lambda_{m,p}^{k,\bar{u}}[\ell] \lambda_{m,q}^{k,\bar{u}}[\ell']}{\big|\zeta_{pq}^{k,\bar{u}}\big|^2 (N-U)}.
\end{align}
Therefore, only when $u=\bar{u}$, $\ddot{\mathbf{V}}_{m}^{k,u}$ is a non-zero matrix, and its $(i+L_{\bar{\rm p}}, j+L_{\bar{\rm p}})$-th sub-block is approximated by
\begin{align}\label{eq:ddotCii}
    \ddot{\mathbf{V}}_{m,ij}^{k,\bar{u}} = \sum_{p,q} \theta_{m,p}^{i} \theta_{m,q}^{j} \ddot{\mathbf{D}}_{m,pq}^{k,\bar{u}} + \theta_{m,p}^{i} \theta_{m,q}^{j} \frac{\ddot{\mathbf{D}}_{m,qp}^{k,\bar{u}} - \zeta_{qp}^{k,\bar{u}}\ddot{\mathbf{D}}_{m,pp}^{k,\bar{u}} - \zeta_{pq}^{k,\bar{u}}\ddot{\mathbf{D}}_{m,qq}^{k,\bar{u}} + \big|\zeta_{pq}^{k,\bar{u}}\big|^{2}\ddot{\mathbf{D}}_{m,pq}^{k,\bar{u}}}{\big|\zeta_{pq}^{k,\bar{u}}\big|^{2} (N-U)}.
\end{align}
By applying \eqref{eq:dotCii}, \eqref{eq:ddotCii} into \eqref{eq:Cmku} and integrating equations \eqref{eq:Emiu}, \eqref{eq:Rmu}, \eqref{eq:Cmku}, the closed-form expression of $\mathcal{R}_{\bar{m}}^{\bar{u}}$ with ZF combiner can be obtained.

\section{Simulation Results}\label{sim}
In this section, simulation results are provided to evaluate the proposed precoder performance and confirm the derived closed-form ergodic achievable rate expression. We consider a circular area, and simulation parameters are listed in Table \ref{tab:sim} unless stated otherwise. Denote the signal-to-noise ratio (SNR) for all APs as $\rho = 1/\sigma_{\eta}^{2}$ as the power of each OQAM symbol is normalized. The results are obtained by Monte Carlo simulation, which are averaged over multiple implementations of channels and user positions. Theoretically, the range of time offset represented by the number of samples for a user to receive AP signals is $[0,\tau_{\rm max}]$, where $\tau_{\rm max}$ is calculated as $\tau_{\rm max} = \frac{2R}{\mathcal{C}T_{\rm s}} \approx 51$.

\begin{table}[h]
    \caption{Parameters setup for simulation}
    \label{tab:sim}
    \centering
    \begin{tabular}{ccc}
        \toprule
        Radius of circular area $R$ & & $1000$ m \\
        Number of APs $K$ & & $8$ \\
        Number of users $U$ & & $4$ \\
        Number of AP antennas $N$ & & $16$ \\
        Number of subcarriers $M$ & & $256$ \\
        Subcarrier spacing & & $30$ kHz \\
        Channel model & & 3GPP-EVA \cite{3GPP} \\
        Large-scale fading $\beta_{k,u}$ & & Three-slope model \cite{HAHE17} \\
        Prototype filter $f[\ell]$ & & PHYDYAS \cite{MB10} \\
        Overlapping factor $\kappa$ & & $4$ \\
        Upsampling factor $C_1$ & & $2$ \\
        Precoder length $L_{\rm p}$ & & $3$ \\
        \bottomrule
    \end{tabular}
\end{table}

\begin{figure}[h]
    \centering
    \includegraphics[width=2.7in]{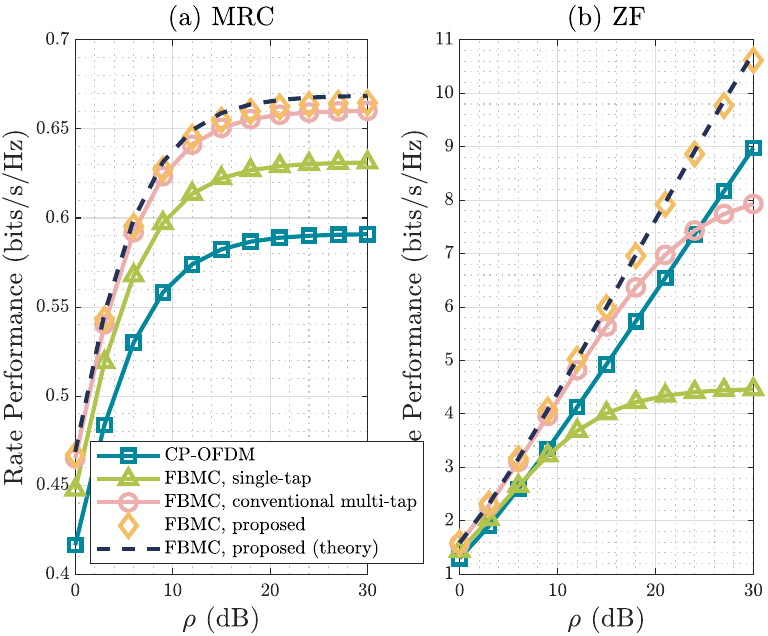}
    \centering
    \caption{Rate performance of different precoding schemes versus $\rho$.}
    \label{fig:rate_SNR}
\end{figure}

Fig. \ref{fig:rate_SNR} shows the rate performance of different precoding schemes of a certain user at a certain subcarrier versus $\rho$, where ``conventional multi-tap'' refers to the scheme with only one $\frac{M}{2}$-fold interpolation and multi-tap precoders performing before the interpolation. For the proposed multiple interpolation scheme, the theoretical rate performance derived in Section \ref{sec:PA} is marked as ``theory'' in the legend. Besides, CP-OFDM based DM-MIMO system with single-tap precoding is also added for comparison, whose rate performance is expressed as
\begin{align}
    \mathcal{R}_{\text{OFDM},\bar{m}}^{\bar{u}} = \frac{M}{M+N_{\rm cp}} \mathbb{E}\left\{\log_2 \left( 1+\gamma_{\text{OFDM},\bar{m}}^{\bar{u}} \right) \right\},
\end{align}
where $N_{\rm cp}$ is the CP length and $\gamma_{\text{OFDM},\bar{m}}^{\bar{u}}$ represents the signal-to-interference-plus-noise ratio (SINR). Since increasing $N_{\rm cp}$ can improve $\gamma_{\text{OFDM},\bar{m}}^{\bar{u}}$ in the case of asynchronous reception \cite{YMDD11} but also reduce the spectral efficiency, we make a trade-off and find an optimal CP length to maximize the ergodic achievable rate through enumeration. We can see that the theoretical results match the numerically simulated results perfectly for both MRC and ZF combiners, which validates the derived closed-form expressions. Compared with conventional multi-tap precoding scheme, our precoding scheme has better performance as the response difference phenomenon has significant impact on the multi-tap precoder design, and the proposed multiple interpolation structure can effectively reduce the impact of the phenomenon. Moreover, FBMC with the proposed scheme outperforms CP-OFDM as the former reduces the impact of asynchronous reception through time-frequency well localized filter bank and well designed precoders instead of long CP.

\begin{figure}[h]
    \centering
    \includegraphics[width=2.7in]{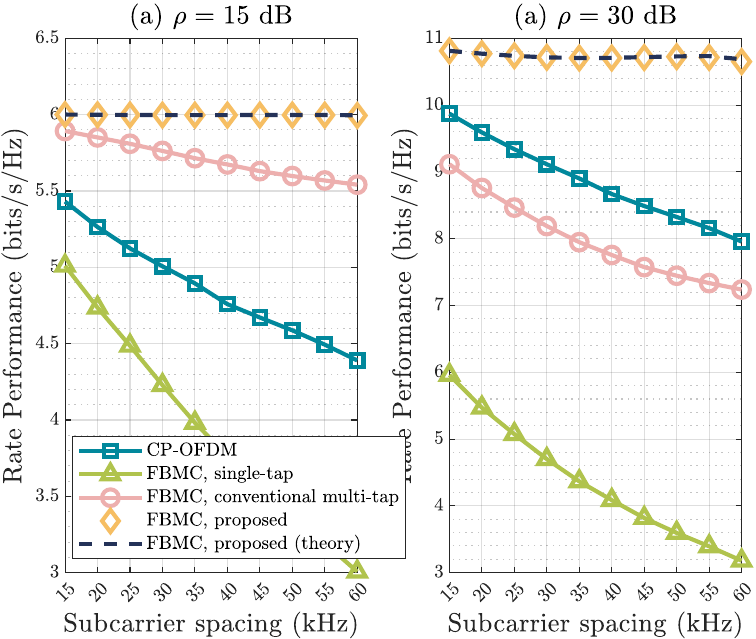}
    \centering
    \caption{Rate performance of different precoding schemes  with ZF combiner versus subcarrier spacing.}
    \label{fig:rate_ss}
\end{figure}

Fig. \ref{fig:rate_ss} shows the rate performance versus the subcarrier spacing, where only ZF combiner is considered due to the poor performance of MRC. We can see that since the time offsets increase with the subcarrier spacing, the achievable rates of other schemes decrease evidently. However, the performance of the proposed scheme can still maintain at a relatively high level, which benefits from the absence of CP for FBMC and the effect of multiple interpolation structure on asynchronous reception and frequency-selective channels. Besides, our derived theoretical rate performance can accurately characterize the numerical results under different subcarrier spacing.

\begin{figure}[h]
    \centering
    \includegraphics[width=2.7in]{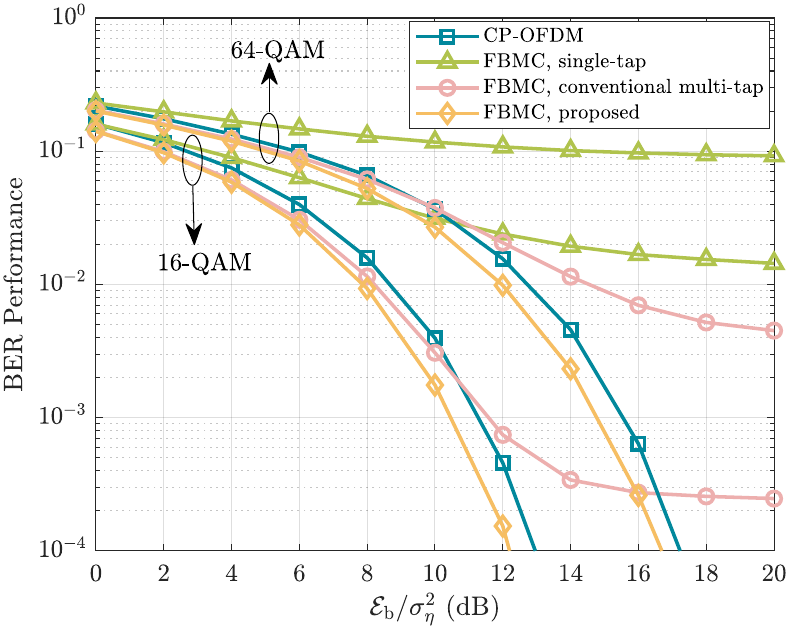}
    \centering
    \caption{BER performance of different precoding schemes with ZF combiner.}
    \label{fig:BER}
\end{figure}

In Fig. \ref{fig:BER}, we compare BER performance of the precoding schemes with ZF combiner, where the CP length is also optimal for CP-OFDM. $\mathcal{E}_{\rm b}$ represents the required power to transmit one bit. It can be seen that FBMC with our proposed scheme has the best performance regardless of modulation order, which illustrates that FBMC is more applicable in cell-free scenarios with asynchronous downlink reception.

\section{Conclusion}
In this work, we developed a practical precoding design scheme for downlink cell-free DM-MIMO FBMC/OQAM with asynchronous reception and highly frequency-selective channels. Besides, we analyzed the rate performance of the proposed scheme, and derived the closed-form expression. Simulation results showed that the theoretical expression could approach numerical results, and FBMC with the proposed scheme had better rate and BER performance compared with CP-OFDM and conventional multi-tap precoding for FBMC in the asynchronous scenario.


\begin{thebibliography}{99}

    \bibitem{HAHE17}
    H. Q. Ngo, A. Ashikhmin, H. Yang, E. G. Larsson, and T. L. Marzetta, ``Cell-free massive MIMO versus small cells,'' \emph{IEEE Trans. Wireless Commun.}, vol. 16, no. 3, pp. 1834-1850, Mar. 2017.

    \bibitem{LLZW21}
    J. Li, M. Liu, P. Zhu, D. Wang, and X. You, ``Impacts of asynchronous reception on cell-free distributed massive MIMO systems,'' \emph{IEEE Trans. Veh. Technol.}, vol. 70, no. 10, pp. 11106-11110, Oct. 2021.

    \bibitem{BFB11}
    B. Farhang-Boroujeny, ``OFDM versus filter bank multicarrier,'' \emph{IEEE Signal Processing Mag.}, vol. 28, no. 3, pp. 92-112, May 2011.

    \bibitem{RNSM17}
    R. Nissel, S. Schwarz, and M. Rupp, ``Filter bank multicarrier modulation schemes for future mobile communications,'' \emph{IEEE J. Sel. Areas Commun.}, vol. 35, no. 8, pp. 1768-1782, Aug. 2017.

    \bibitem{YMDD11}
    Y. Medjahdi, M. Terre, D. L. Ruyet, D. Roviras, and A. Dziri, ``Performance analysis in the downlink of asynchronous OFDM/FBMC based multi-cellular networks,'' \emph{IEEE Trans. Wireless Commun.}, vol. 10, no. 8, pp. 2630-2639, Aug. 2011.

    \bibitem{DGXM16}
    D. Gregoratti and X. Mestre, ``Uplink FBMC/OQAM-based multiple access channel: Distortion analysis under strong frequency selectivity,'' \emph{IEEE Trans. Signal Process.}, vol. 64, no. 16, pp. 4260-4272, Aug. 2016.

    \bibitem{SYDA21}
    S. Mahama, Y. J. Harbi, D. Grace, and A. G. Burr, ``Multi-user interference cancellation for uplink FBMC-based multiple access channel,'' \emph{IEEE Commun. Lett.}, vol. 25, no. 8, pp. 2733-2737, Aug. 2021.


    \bibitem{QDZW23}
    Y. Qi, J. Dang, Z. Zhang, L. Wu, and Y. Wu, ``Efficient channel equalization and performance analysis for uplink FBMC/OQAM-based massive MIMO systems,'' \emph{IEEE Trans. Veh. Technol.}, early access.

    \bibitem{HABF22}
    H. Hosseiny, A. Farhang, and B. Farhang-Boroujeny, ``Downlink precoding for FBMC-based massive MIMO with imperfect channel reciprocity,'' in \emph{Proc. of IEEE Int. Conf. Commun. (ICC)}, May 16-20, Seoul, Korea, 2022, pp. 1324-1329.

    \bibitem{PPV93}
    P. P. Vaidyanathan, \emph{Multirate Systems and Filter Banks}, Prentice Hall PTR, Englewood Cliffs, 1993.

    \bibitem{ZJWZ14}
    Q. Zhang, S. Jin, K. -K. Wong, H. Zhu, and M. Matthaiou, ``Power scaling of uplink massive MIMO systems with arbitrary-rank channel means,'' \emph{IEEE J. Sel. Topics Signal Process.}, vol. 8, no. 5, pp. 966-981, Oct. 2014.

    \bibitem{PRKV19}
    P. Singh, R. Budhiraja, and K. Vasudevan, ``SER analysis of MMSE combining for MIMO FBMC-OQAM systems with imperfect CSI,'' \emph{IEEE Commun. Lett.}, vol. 23, no. 2, pp. 226-229, 2019.

    \bibitem{DKAK11}
    D. K. Nagar and A. K. Gupta, ``Expectations of functions of complex Wishart,'' \emph{Acta Appl. Math.}, vol. 113, no. 3, pp. 265-288, Mar. 2011.

    \bibitem{3GPP}
    \emph{Technical Specification Group Radio Access Network; Evolved Universal Terrestrial Radio Access (E-UTRA); User Equipment (UE) Radio Transmission and Reception}, document TS 36.101, 3GPP, Dec. 2007. [Online]. Available: http://www.3gpp.org

    \bibitem{MB10}
    M. Bellanger \emph{et al}., ``FBMC physical layer: A primer,'' \emph{PHYDYAS}, vol. 25, no. 4, 2010.

\end{thebibliography}
\end{document}